\documentclass[reprint,amsmath,amssymb,aps]{revtex4-2}

\usepackage{graphicx}
\usepackage{dcolumn}
\usepackage{bm}
\usepackage{comment}
\usepackage{xcolor}
\usepackage[caption=false]{subfig}
\usepackage[resetlabels,labeled]{multibib}
\newcites{SM}{SM References}

\begin{document}

\preprint{APS/123-QED}

\title{$P$-wave Efimov physics implications at unitarity}

\author{Yu-Hsin Chen}
\email{chen2662@purdue.edu}
\affiliation{Department of Physics and Astronomy, Purdue University, West Lafayette, Indiana 47907 USA}
\author{Chris H. Greene}
\email{chgreene@purdue.edu}
\affiliation{Department of Physics and Astronomy, Purdue University, West Lafayette, Indiana 47907 USA}
\affiliation{Purdue Quantum Science and Engineering Institute, Purdue University, West Lafayette, Indiana 47907 USA}

\date{\today}

\begin{abstract}
Equal mass fermionic trimers with two different spin components near the unitary limit are shown to possess a universal van der Waals bound or resonance state near $s$-wave unitarity, when $p$-wave interactions are included between the particles with equal spin. Our treatment uses a single-channel Lennard-Jones interaction with long range two-body van der Waals potentials. While it is well-known that there is no true Efimov effect that would produce an infinite number of bound states in the unitary limit, we demonstrate that another type of universality emerges for the symmetry $L^{\Pi}=1^{-}$.  The universality is a remnant of Efimov physics that exists in this system at $p$-wave unitarity, and it leads to modified threshold and scaling laws in that limit.  Application of our model to the system of three lithium atoms studied experimentally by Du, Zhang, and Thomas [Phys. Rev. Lett. {\bf 102}, 250402 (2009)] yields a detailed interpretation of their measured three-body recombination loss rates.
\end{abstract}

\maketitle
Dilute bosonic and fermionic ultracold gases have benefited tremendously in recent decades from the ability to accurately tune $s$-wave scattering lengths or $p$-wave scattering volumes at two-body Fano-Feshbach resonances\cite{inouye1998observation}. In the most-studied scenario, the $s$-wave scattering length ($a_s$) between each pair of atoms in a system of three identical bosons diverges to negative infinity and produces the infinite series of trimer bound or resonance state energies in the now-familiar Efimov effect\cite{efimov1970energy,efimov1970SJNP,braaten2006universality,wang2013ultracold,naidonEfimovPhysicsReview2017,d2018few}.  
From a hyperspherical coordinate perspective, the Efimov effect occurs because, at two-body unitarity,  the long-range adiabatic hyperradial potential curve is reduced from a repulsive to an attractive asymptotic potential $W(R) \rightarrow -(s_0^2+\frac{1}{4})\hbar^2/(2\mu R^2)$ with $s_0 \approx 1$.  Here $R$ is the hyperspherical radius and $\mu$ is the three-body reduced mass\cite{suno2002three}. 
Several experimental groups have since demonstrated that the Efimov effect can be observed in helium trimers\cite{kunitski2015observation} and in some alkali atom trimers\cite{kraemer2006evidence,berninger2011universality,zaccanti2009observation,pollack2009universality,gross2009observation,ottenstein2008collisional,wild2012measurements}. Also, for three identical bosons with orbital angular momentum and parity $L^{\Pi}=0^{+}$, the scaling of three-body recombination has been predicted and observed \cite{esry1999recombination,braaten2006universality} to vary approximately in proportion to $|a_{s}|^{4}$ for negative values of $a_s$ far from resonances.

For trimers consisting of single-component fermions, a $p$-wave Efimov effect in the symmetry $L^{\Pi}=1^{-}$ was initially predicted by Macek and Sternberg \textit{et al}\cite{macek2006properties}. However, that prediction was disproven by effective field theory; that prediction of a $p$-wave Efimov effect in zero-range theory turned out to be unphysical since it led to energy eigenstates having negative probability\cite{braaten2006universality,nishida2012impossibility}. In the two-component fermionic equal mass trimers, there are no known symmetries having an Efimov effect at $s$-wave unitarity. For the unequal mass system of two spin-up fermions and one spin-down fermion, the Efimov effect occurs when the mass ratio of spin-up fermions to a spin-down fermion is greater than $\approx 13.6$\cite{braaten2012renormalization}. Braaten \textit{et al} also found an Efimov effect at $p$-wave unitarity when there is a strong $p$-wave interaction between the third equal mass particle and each of the two identical fermions, for symmetries $L^{\Pi}=0^{+},1^{+},1^{-}$ and $2^{+}$ or two identical bosons with $L^{\Pi}=1^{+}$\cite{braaten2012renormalization}. Nevertheless, these equal mass cases of a predicted Efimov effect have all been proven by Nishida \textit{et al.}\cite{nishida2012impossibility} to be unphysical.

Other cases of unequal mass universal trimers have been predicted, as in \cite{kartavtsev2007low}.
Moreover, in a recent study, Naidon {\it et al.} found a universal trimer state that occurs when the heavy/light mass ratio is smaller than 8.2, all the way down to the equal mass regime, when the $p$-wave interaction between two spin-up fermion\cite{naidon2021shallow} is tuned near unitarity. The scaling law of two-component equal mass fermionic three-body recombination rate was predicted by D'Incao {\it et al} to be $|a_{s}|^{2.455}$ for $a_{s}<0$\cite{d2005scattering}, as well as by D'Incao {\it et al} and Petrov to be $a_{s}^{6}$ for $a_{s}>0$\cite{d2005scattering,petrov2003three}. For the negative $a_s$ side, the scaling law of three-body recombination ($K_3$) disagrees significantly with the experimental result that has observed $K_{3}\propto |a_{s}|^{0.79 \pm 0.14}$ for $a_s < 0$\cite{du2009inelastic}. One point about this theory-experiment discrepancy, which is relevant to considerations discussed below in the present study, is that those previous theoretical predictions neglect the interaction between the two spin-up fermions, which might potentially be responsible for the discrepancy in the threshold law compared to the experimental results. 

The present letter predicts the existence of one $p$-wave universal trimer state for two-component equal mass ($m_{\uparrow}=m_{\downarrow}$) fermionic trimers having the $s$-wave scattering length approaching infinity between spin-up and spin-down fermions and having the $p$-wave scattering volume ($V_p$) at the unitary limit, for the symmetry $L^{\Pi}=1^{-}$. Evidence for the universality of this trimer state has emerged from our tests of different $s$- and/or $p$-wave poles of two-body Lennard-Jones potentials, in addition to various other two-body potentials, e.g., Gaussian or Gaussian-type potentials\cite{higgins2022three,chen2022efimov}. The peak of the three-body recombination rate can indicate the starting point where the trimer state becomes bound. The scaling law for the three-body recombination rate as a function of the $p$-wave scattering volume $V_p$ is modified when the opposite-spin fermion interactions are at or near $s$-wave unitarity. For a two-component Fermi gas trimer, especially the two spin-up and one spin-down fermion case, the symmetry $L^{\Pi}=1^{-}$ dominates near the three-body dissociation threshold, because the recombination rate linearly on the collision energy $E$\cite{esry2001threshold}. However, the $K_{3}\propto E$ no longer applies at the $s$-wave or $p$-wave unitary limit for the opposite spin fermions, or if the two spin-up fermions interact at $p$-wave unitarity. The calculations presented here also overcome a longstanding inability of theory to understand the recombination rate measured in an experiment\cite{du2009inelastic} carried out at very large $s$-wave scattering lengths. 

The adiabatic hyperspherical representation has a strong track record in describing few-body interactions and collisional phenomena\cite{Rittenhouse-2011JPB,Greene2017RMP,wang2013ultracold,d2018few}, and is used here to analyze the three-body quantum problem. For our present studies, the two spin-up fermions interacting with one spin-down fermion (or any other equal mass particle) with symmetry $L^{\Pi}=1^{-}$, where $\Pi$ is the total parity in the system. The three-body Schr\"{o}dinger equation is rewritten using  modified Smith-Whitten hyperspherical  coordinates\cite{whitten1968symmetric,johnson1980hyperspherical,kendrick1999hyperspherical,suno2002three}:
\begin{equation}\label{eq:Schrodingereq}
	 \left[-\frac{\hbar^2}{2\mu}\frac{d^{2}}{d R^{2}}+W_{\nu}(R)\right]F_{\nu}(R)+\sum_{\nu\neq\nu'}W_{\nu\nu'}F_{\nu'}(R)=E F_{\nu}(R)
\end{equation}
Here $R$ is the hyperspherical radius, $\mu = m/\sqrt{3}$ is the three-body reduced mass for three equal particles with mass $m$, $W_{\nu}(R)$ is the effective {\it adiabatic potential} in channel $\nu$, $F_{\nu}(R)$ is the hyperradial wavefunction and $W_{\nu\nu'}(R)$ is the nonadiabatic coupling. The full interaction potential energy $V$ is taken here to be a sum of the two-body potentials, i.e., $V = v_3(r_{12})+v_1(r_{23})+v_2(r_{31})$,
where the $r_{ij}$ are the interparticle distances. The two-body potential is the following, e.g., in the case of the Lennard-Jones potential\cite{wang2012origin}
\begin{equation}
	v_i(r)=-\frac{C_{6}}{r^{6}}\left(1-\frac{\lambda_n^{6}}{r^{6}}\right)
\end{equation}
In this letter, in our chosen set of van der Waals units, the $C_6$ coefficient is set at $16 \, r_{\text{vdW}}^{6}E_{\text{vdW}}$ where $r_{\text{vdW}}$ is the van der Waals length $r_{\text{vdW}}\equiv(m C_6/\hbar^2)^{1/4}/2$, ($m/2$ is two-body reduced mass here) and $E_{\text{vdW}}$ is the van der Waals energy unit, $E_{\text{vdW}}\equiv \hbar^2/(2\mu r_{\text{vdW}}^2)$. The parameter $\lambda_n$ can be adjusted to produce any desired $s$-wave scattering length or $p$-wave scattering volume for a chosen pair of fermions. Here the two-body $s$-wave scattering length and $p$-wave scattering volume can be generally written as
\begin{equation}
    k^{2L+1} \cot(\delta_L) = -1/{a_L}^{2L+1}+\frac{1}{2} r_L k^2 
\end{equation}
where the $a_0(\equiv a_s)$ is the $s$-wave scattering length, the $a_1(\equiv a_p)$ is the $p$-wave scattering length, the $\delta_{0}(k)$ is $s$-wave scattering phase shift, the $\delta_{1}(k)$ is $p$-wave scattering phase shift, the $r_0$ is the $s$-wave effective range, the $r_1$ is the $p$-wave effective range in units of inverse length, and $k$ is the wave number.  

\begin{figure}[h]
\centering
\includegraphics[width=8cm]{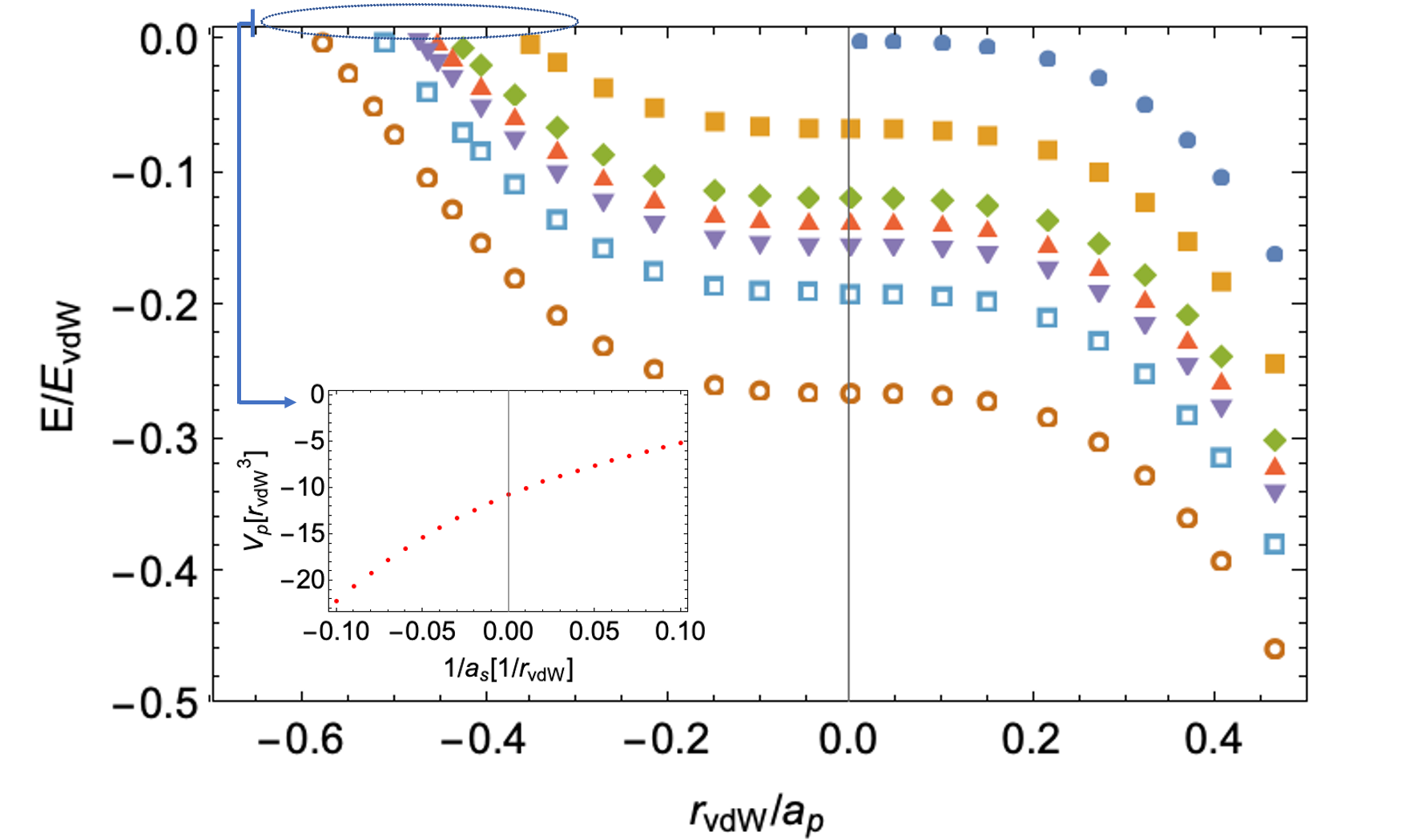}
\caption{(color online). Shown are $p$-wave universal trimer state energies for two spin-up and one spin-down fermions system ($\uparrow\downarrow\uparrow$) with $L^{\Pi}=1^{-}$ versus the inverse of $p$-wave scattering length  ($a_p \equiv V_{p}^{1/3}$). 
Circles (solid blue) show the two-body $p$-wave bound state.  The squares, diamonds, triangles, inverted-triangles, open squares and open circles represent respectively the trimer energies obtained using different fixed interactions (scattering lengths) between the spin-up and spin-down fermions ($\uparrow\downarrow$) plotted here as functions of the $p$-wave $a_p$ between the two spin-up fermions ($\uparrow\uparrow$).   The respective $s$-wave values of $a_s$ in the order listed above for the different symbols are $a_{s}=-10\,r_{\text{vdW}}$, $a_{s}=-50\,r_{\text{vdW}}$, $a_{s}=\infty \,r_{\text{vdW}}$, $a_{s}=50\,r_{\text{vdW}}$, $a_{s}=20\,r_{\text{vdW}}$ and $a_{s}=10\,r_{\text{vdW}}$. The inset figure shows the starting value of $a_p$ where two spin-up and one spin-down fermion first form the trimer state at zero energy for the corresponding value of the $s$-wave scattering length between opposite spin state fermions.}
\label{fig:spsVp}
\end{figure}

The asymptotic effective adiabatic potentials in the 3-body continuum are accurately characterized at $R\rightarrow\infty$ as
\begin{equation}\label{eq:wasym}
	W_{\nu}(R) \rightarrow \frac{\hbar^2 l_{e}(l_{e}+1)}{2\mu R^{2}}.
\end{equation}
where $l_e$ controls the effective angular momentum barrier of the three free asymptotic particles in the large hyperradius, $R\rightarrow\infty$. The $l_e$ value also determines the
scaling law of the three-body recombination rate and squared scattering matrix element, through the Wigner threshold law $|S_{j\leftarrow i}^{L\Pi}|^{2} \propto {k_i}^{2l_{e,i}+1}$ $R$-matrix propagation is used to solve the radial Eq.(\ref{eq:Schrodingereq}), after which the $S$-matrix and the three-body recombination rate ($K_{3}^{L\Pi}$) can be computed using
\begin{equation}\label{eq:recom}
    K_{3}^{L\Pi}=\sum_{L,\Pi}\sum_{i,j}\frac{32\hbar N! (2L+1)}{\mu k^4}|S_{j\leftarrow i}^{L\Pi}|^{2}.
\end{equation}
Here $N$ is the number of particles in the trimer that are identical, $k=\sqrt{2\mu E/\hbar^{2}}$ is the hyperspherical wave number, $E$ is the three-body collision energy, and $i$ and $j$ label the incident (three-body continuum) and outgoing (three-body recombination) channels attached to two-body energies\cite{mehta2009general,wang2011numerical}. In our treatment, the situation with two spin-up and one spin-down fermion can be viewed as two identical fermions plus a third atom of equal mass, in which case $N=2$ in Eq.(\ref{eq:recom}). According to the Wigner threshold law, the three-body recombination rate is a power-law function of the incident wave number $k=\sqrt{2\mu E/\hbar^{2}}$ at ultracold energy, namely
\begin{equation}\label{eq:recom2}
    K_{3}^{L\Pi}\propto k^{2l_e-3}
\end{equation}
Therefore, the coefficient of $1/(2\mu R^2)$ in the asymptotically lowest continuum effective adiabatic potential plays a key role in the behavior of the low energy three-body recombination rate. 

Our explorations demonstrate the existence of a $p$-wave universal trimer for three equal mass fermionic atoms at the $s$-wave unitary limit, which emerges when the interaction between the two spin-up atoms is made attractive, specifically for the symmetry $L^{\Pi}=1^{-}$. Even if these fermions do not have a divergent $s$-wave scattering length ($a_s$), however, the $p$-wave universal trimer state still exists. Fig.\ref{fig:spsVp} plots the $p$-wave trimer state energy versus the $p$-wave scattering length ($a_p \equiv V_{p}^{1/3}$) between the identical fermions; this was calculated by including 30 coupled continuum channel potential curves. Different symbols represent the different $s$-wave scattering lengths between the opposite spin state fermions. As one increases the attraction in the $p$-wave potential between spin-polarized fermion, it is seen that the trimer can be created not only at the $s$-wave unitary limit for the opposite-spin fermion interactions, but even at small positive or negative values of the $s$-wave scattering length. These can be viewed as an extension of predictions for the KM trimer\cite{kartavtsev2007low}, i.e., for arbitrary $s$-wave interactions between different spin state fermions as one enhances the $p$-wave attraction between the same spin state fermion in the overall symmetry $L^{\Pi}=1^{-}$. 
In the case where the unequal spin interactions are fixed at $s$-wave unitarity, and the interactions between like fermions are at $p$-wave unitarity, the universal trimer energy is computed here to equal $E=-0.1355\,E_{\text{vdW}}$ where the two-body $s$-wave and $p$-wave effective ranges are $r_s\approx 2.782\,r_{\text{vdW}}$ and $r_p\approx -1.727\, r_{\text{vdW}}^{-1}$, respectively. 
The critical points where the universal trimer state reaches zero energy and causes a recombination resonance can be determined from the axis intercepts in Fig.\ref{fig:spsVp}. The inset of Fig.\ref{fig:spsVp} shows those critical values of $V_p$ as a function of $a_s$, as the $s$-wave scattering length ranges from $-10\,r_{\text{vdW}}$ to $\infty$ and on to $10 \,r_{\text{vdW}}$. For this $a_s$ regime, there is no additional $p$-wave two-body resonance, nor has a $p$-wave Feshbach molecule been created. The inset figure suggests how experiment can find the recombination resonance associated with the universal trimer state, when the different spin fermion interactions differ from the $s$-wave unitary limit.

\begin{figure}[h]
\includegraphics[width=8cm]{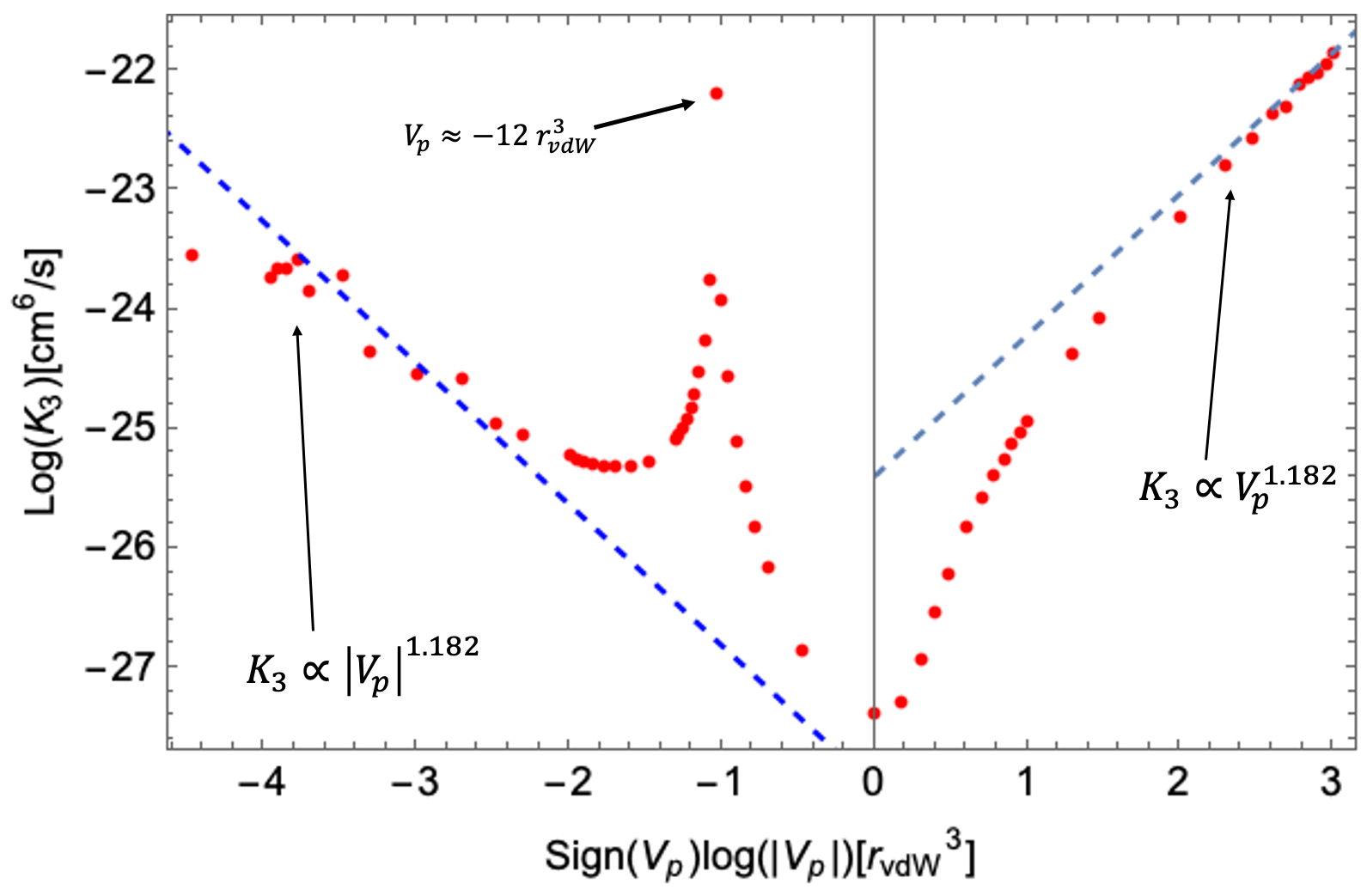}
\caption{(color online). Our numerically computed three-body recombination rate is shown as a function of the $p$-wave scattering volume $V_p$ for the symmetry $L^{\Pi}=1^{-}$. This plot represents the three ${}^{6}_{}{\text Li}$ atoms with two spin-up and one spin-down state ($\uparrow\downarrow\uparrow$) at the temperature $T \approx 150 \, nK$. The peak of the recombination rate at $V_p \approx -12 \, r_{\text{vdW}}^{3} $ close to the previously discussed
starting point where the universal trimer first becomes bound, near the $1^{\text{st}}$ $p$-wave pole of two-body Lennard-Jones potential.  Moreover, the dashed line asymptote shows the expected scaling law of the recombination rate for the computed value of $l_e$, namely for this symmetry $K_3 \propto |V_{p}|^{1.182}$. The x- and y-axis are both logarithmic scales, base 10.} 
\label{fig:recom}
\end{figure}

Measuring the three-body recombination rate gives a way to find values of the two-body scattering parameters where the universal trimer state hits zero energy.  Consider next the situation where unlike spins have their interaction fixed at $s$-wave unitarity, as the $p$-wave scattering volume $V_p$ is varied between the same spin fermions. Fig.\ref{fig:recom} plots the three-body recombination rate versus  $V_p$ for two spin-up and one spin-down ($\uparrow\downarrow\uparrow$) ${}^{6}_{}{\text {Li}}$ atom with the symmetry $L^{\Pi}=1^{-}$. The resonant peak of the recombination rate corresponds to the creation of the universal trimer state, predicted here to occur when the $p$-wave scattering volume is $V_p\approx -12 \, r_{\text{vdW}}^{3}$. This recombination rate was calculated by including 6 atom-dimer channels for recombination channels and 14 continuum channels. This $V_p$ is close to its value predicted in our inset of Fig.\ref{fig:spsVp}, $V_p\approx -11.06 \, r_{\text{vdW}}^{3}$ (including 14 continuum channels and no atom-dimer channels).
Note the dramatic dependence on $V_p$: the peak value of $K_3$ being around five orders of magnitude higher than the $K_3$ value at zero $p$-wave scattering volume. The asymptotic behavior of the recombination rate $K_3$, for $|V_p| \gg r_{vdW}^3$, has been modified to $|V_p|^{(2l_e+1)/3}\rightarrow |V_p|^{1.181}$ since the value of $l_e$ controlling the large-$R$ adiabatic potential is modified to $l_e=1.272$ by Efimov physics for the two-component fermion system in the $s$-wave unitary limit. 
The Efimov physics modification\cite{chen2022efimov,higgins2022three}, for this two-component Fermi trimer at $s$-wave unitarity in the symmetry $L^{\Pi}=1^{-}$, can be obtained by solving the transcendental equation for zero-range interactions, as in Refs. \cite{werner2006unitary,blume2007universal}. 
We confirm numerically that the lowest $l_e$ value, which sets the long-range barrier of the three-body effective potential, controls both the Wigner threshold law Eq.[\ref{eq:recom2}] and the scaling of the three-body recombination rate with the $p$-wave scattering volume. Hence, by fixing the different spin interaction at $s$-wave unitarity (infinite $a_s$), not only can this previously unobserved universal trimer state be found, but also the scaling law of the three-body loss rate as a function of $V_p$ can be tested by tuning the $p$-wave interaction between the pair of spin-up fermionic atoms. 

\begin{figure}[h]
\includegraphics[width=9cm]{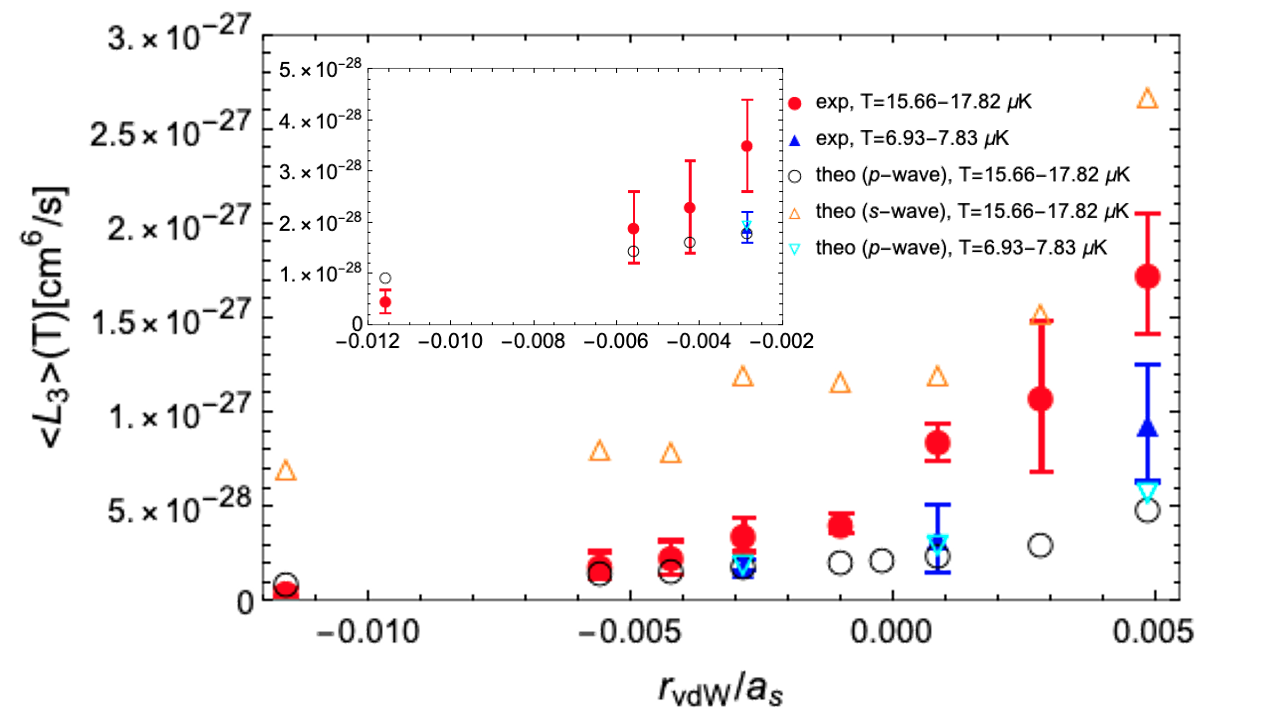}
\caption{(color online). Comparison of the theoretical thermally-averaged three-body loss rate for ${}^{6}_{}{\text{Li}}$ with the Du {\it et al.} experiment\cite{du2009inelastic} at large scattering length $|a_s|$, using the two spin-up and one spin-down ($\uparrow\downarrow\uparrow$) model with trimer orbital angular momentum $L^{\Pi}=1^-$ and the temperature range is from $T\approx 15.66\, \mu K$ to $T\approx 17.82\,
\mu K$. The filled circles (red) represent the experimental data from $T\approx 15.66\, \mu K$ to $T\approx 17.82\, \mu K$, the filled triangles (blue) are the experimental data at lower temperature from $T\approx 6.93\, \mu K$ to $T\approx 7.83\, \mu K$. The open circles (black), open triangles (orange) and inverted-triangles (cyan) are our numerical calculations of the thermally-averaged three-body loss rate for two different Hamiltonians with various, as is explained in the text.}
\label{fig:John}
\end{figure}

The thermally averaged three-body recombination rate is written (after correcting a typo in Eq.(4) of Ref.\cite{suno2003recombination}) as
\begin{equation}
    \langle K_3 \rangle(T)=\frac{1}{2(k_B T)^3}\int K_3(E)E^2 e^{-E/(k_{B}T)}dE.
\end{equation}
Our computed recombination rate shows generally good agreement with the ternary loss rate measurement carried out by the group of Thomas\cite{du2009inelastic} for a two-component gas of fermionic ${}^{6}_{}{\text {Li}}$.  Fig.\ref{fig:John} compares theory and experiment in a plot of the thermally averaged {\it atom-loss} rate  $\langle L_3 \rangle(T)$. This quantity is related to the event loss rate $K_3$ through the equation $\langle L_3 \rangle(T)=3\langle K_3 \rangle(T)/2$ \cite{esry1999recombination}, where the brackets denote thermal averaging. The computed results in Fig.\ref{fig:John} are for the recombination of two spin-up and one spin-down  ${}^{6}_{}{\text {Li}}$ atoms in the symmetry $L^{\Pi}=1^{-}$, and are shown as a function of $r_{\text{vdW}}/a_s$, which controls the interaction of opposite spin state fermions. Moreover, the $p$-wave scattering volume between two spin-up fermion has been fixed at $V_p\approx -1.8\, r_{\text{vdW}}^{3}$ (This value of $V_p$ was obtained in multichannel calculations carried out by Yimeng Wang in Ref.\cite{wang2022multi}). For two ${}^{6}_{}{\text {Li}}$ atoms, the van der Waals length is equal to $r_{\text{\text{vdW}}}=31.26\,a_{B}$ and the van der Waals energy is $E_{\text{vdW}}/k_{B}=29.47\,m K$, where $a_{B}$ and $k_{B}$ are the Bohr radius and the Boltzmann constant, respectively\cite{chin2010feshbach}. 
The open circles (black),  open inverse-triangle (cyan) and open triangles (orange) points show the thermally averaged three-body loss rate $ \langle K_3 \rangle$, for the same temperature range probed experimentally. The open circles were calculated using the three-body Born-Oppenheimer potential curves obtained for a Hamiltonian that includes deep $p$- and $f$-wave atom-dimer recombination channels only, and the inverted-triangles were obtained using the same potential curves but at lower temperature to make comparison with experimental data. The open triangles were calculated instead for a Hamiltonian that possesses deep atom-dimer recombination channels, from $s$-wave to $g$-wave. 

More details about these calculations and the Born-Oppenheimer potential curves are shown in the Supplementary Material. Our study concentrates on the recombination into deep atom-dimer channels for the recombination process; recombination into the shallow $s$-wave atom-dimer channel at large positive values of the $s$-wave scattering length is omitted from the theoretical results shown in Fig.\ref{fig:John}, because of the extremely small binding energy of those universal dimers for the positive scattering lengths used in that experiment, with binding energies smaller than the gas temperature. If recombination into the weakly-bound universal $s$-wave dimer is included, the rate would be several orders of magnitude higher than the largest experimental rate shown in Fig.\ref{fig:John}.  
In Fig.\ref{fig:John}, the (red) circles represent the experimental three-body loss rate coefficient as a function of the $s$-wave scattering length ($a_s$) which is taken from Table.I  of Ref.\cite{du2009inelastic}, with $a_s$ rescaled into units of the van der Waals length. The triangles (blue) show the experimental three-body atom loss rate at lower temperatures. 

In summary, the Fermi gas with equal mass atoms in two spin components has been shown to support a universal trimer that can be created by tuning the $s$-wave and $p$-wave interaction simultaneously, and this trimer produces an observable resonance in the three-body recombination rate. Moreover, the scaling law of the three-body recombination rate as a function of $p$-wave scattering volume is shown to be modified when the pair of spin-up and spin-down fermion near interacts at the $s$-wave unitary limit. The three-body inelastic collision rates have been computed as a function of the $s$-wave scattering length, with the interaction between the two equal spin fermionic lithium atoms included, and it provides a reasonably complete interpretation of the recombination rates measured by Du {\it et al.}. 

\section*{Acknowledgement}
This work was supported in part by NSF Grant Award Numbers 1912350 and 2207977. We thank Jia Wang and Jose D'Incao for sharing their computer programs.

\bibliography{apssamp}

\section*{Supplementary Material}

\setcounter{equation}{0}
\setcounter{figure}{0}
\setcounter{table}{0}
\makeatletter
\renewcommand{\theequation}{S.\arabic{equation}}
\renewcommand{\thefigure}{S\arabic{figure}}
\renewcommand{\thesubsection}{\Alph{subsection}}

\subsection{Born-Oppenheimer potential curves}
\begin{figure}[ht]
\subfloat[\label{fig:swave86}]
{\includegraphics[width=8cm]{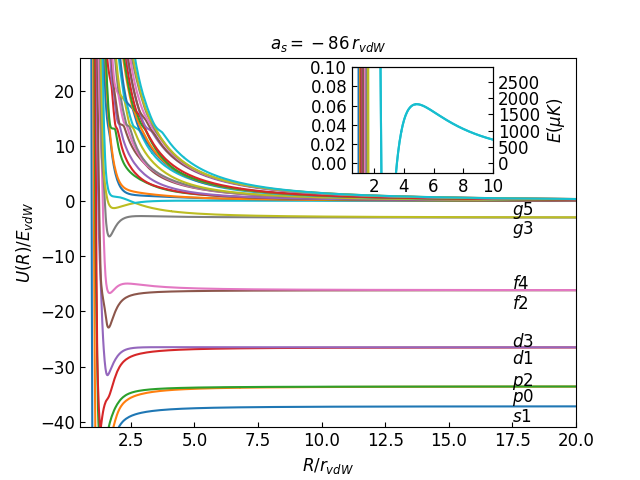}}\centering
\subfloat[\label{fig:swave205}]
{\includegraphics[width=8cm]{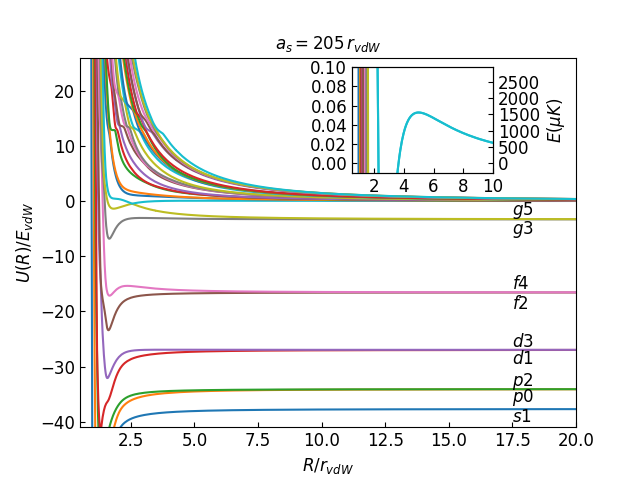}}\\
\subfloat[\label{fig:pwave86}]
{\includegraphics[width=8cm]{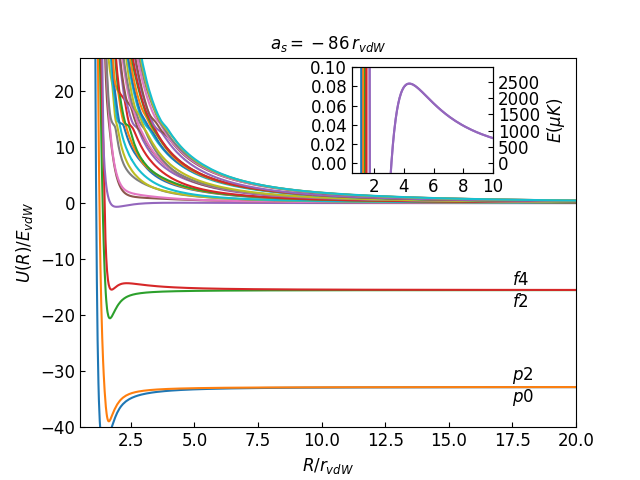}}\centering
\subfloat[\label{fig:pwave205}]
{\includegraphics[width=8cm]{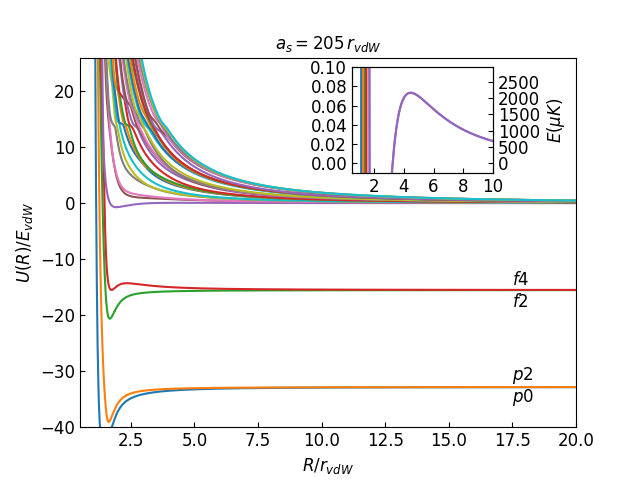}}
\caption{(color online). Three-body Born-Oppenheimer potential curves for two spin-up and one spin-down fermion ($\uparrow\downarrow\uparrow$) with total angular momentum $L^{\Pi}=1^{-}$. The letter represents the angular momentum quantum number $l$ of the dimer, and the number labels the angular momentum between the third atom and dimer $l'$. (a) and (b), the $a_s$ are chosen near the $2^{\text{nd}}$ $s$-wave pole with one deep $s$-wave bound state,
and $V_p=-1.8\,r_{\text{vdW}}^{3}$ with no $p$-wave bound state. 
In (c) and (d),the $a_s$ are chosen near or at the $1^{\text{st}}$ $s$-wave pole with no deep opposite spin dimers, and the potential producing $V_p=-1.8\,r_{\text{vdW}}^{3}$ is chosen to have deep $p$- and $f$-wave bound states that are enable recombination into deep spin-polarized dimers. } 
\label{fig:spwave}
\end{figure}

The lowest several hyperspherical Born-Oppenheimer potential curves for the three-fermion system with symmetry $L^{\Pi}=1^{-}$ are shown in Fig.\ref{fig:spwave}, which enables an interpretation of the difference between the three-body loss rates computed using the two different Hamiltonian described in relation to Fig.3 of the main text. The long-range effective adiabatic potential curve representing the highest atom-dimer channel can be represented asymptotically as
\begin{equation}\label{eq:atomdimer}
    W_{\nu}(R)\xrightarrow[{R\rightarrow\infty}]{} U_{\nu}(R)=E_{\nu l}+\frac{\hbar^2 l'(l'+1)}{2\mu R^2}.
\end{equation}
Here $E_{\nu l}$ is the rovibrational dimer energy, $l$ represents the dimer angular momentum, and $l'$ is the angular momentum of the third particle relative to the dimer. For the three-body hyperspherical calculations shown in Figs.(\ref{fig:swave86}) and (\ref{fig:swave205}), the $s$-wave 2-body potential depth has been chosen to yield the stated values of the scattering lengths ($a_s$) between spin-up and spin-down fermions, in the vicinity of the $2^{\text{nd}}$ $s$-wave pole (i.e., there exists a single deep $s$-wave dimer (in addition to deep $p,d,f$ and $g$ dimers), plus a weakly-bound $s$-wave dimer when $a_s>0$), while the $p$-wave scattering volume between two spin-up fermions was set at $V_{p}=-1.8\,r_{\text{vdW}}^{3}$ (no $p$-wave dimer exists in this 2-body potential). The insets show that the potential barrier exhibits a local maximum (at $R\approx 5\, r_{\text{vdW}}$) in the entrance recombination channel, and tunneling through that barrier to reach smaller hyperradii plays a key role in determining the experimental three-fermion recombination loss rate into an atom and a deep dimer. The barrier decreases gradually as the $s$-wave scattering length is decreased from a small negative value to $-\infty$ and as $a_s$ continues to decrease from $+\infty$ to small positive values ($a_s=-86 \rightarrow \infty  \rightarrow 205\,r_{\text{vdW}}$), i.e., as the interaction potential between opposite spin fermions gets increasingly attractive. Since three-body recombination into deep dimers in this low energy range requires the system to tunnel through that barrier, the gradual decrease of that hyperradial barrier height, as the opposite spin dimer interaction gets more attractive, produces an enhancement of the partial three-body recombination rate. 

Similar reasoning applies to the other two figures Fig.(\ref{fig:pwave86}) and (\ref{fig:pwave205}), the only differences being that: {\it (i)} the $s$-wave interaction between opposite spin state fermion has been chosen near the $1^{\text{st}}$ $s$-wave pole and  {\it (ii)}  the $p$-wave interaction between same spin state fermions is still fixed at $V_{p}=-1.8\,r_{\text{vdW}}^{3}$ but with a deeper vdW potential that supports both deep $p$-wave and $f$-wave bound states. Interestingly, one sees that the computed hyperradial barrier heights in the entrance channels of Fig.(\ref{fig:pwave86}) and (\ref{fig:pwave205}) are about 30\% higher than those plotted in Figs.(\ref{fig:swave86}) and (\ref{fig:swave205}). 

Observe that the open circle recombination rate calculation shown in Fig.3, involves an $s$-wave scattering length $a_s$ between opposite spin state fermions that was computed at or near the $1^{\text{st}}$ $s$-wave pole and where only the two spin-up fermion can form deep dimers of $p$-wave and $f$-wave angular momentum. In this case, the more weakly-bound $f$-wave deep dimer partial recombination rates are computed to be slightly higher than the $p$-wave partial recombination rates.  The relevant hyperradial Born-Oppenheimer potentials for those cases, with $a_{s}=-86\,r_{\text{vdW}}$ and $a_{s}=205\,r_{\text{vdW}}$, are those displayed in Figs.(\ref{fig:pwave86}) and Fig.(\ref{fig:pwave205}), respectively. 

\subsection{Wigner threshold law}

\begin{figure}[ht]
\includegraphics[width=9cm]{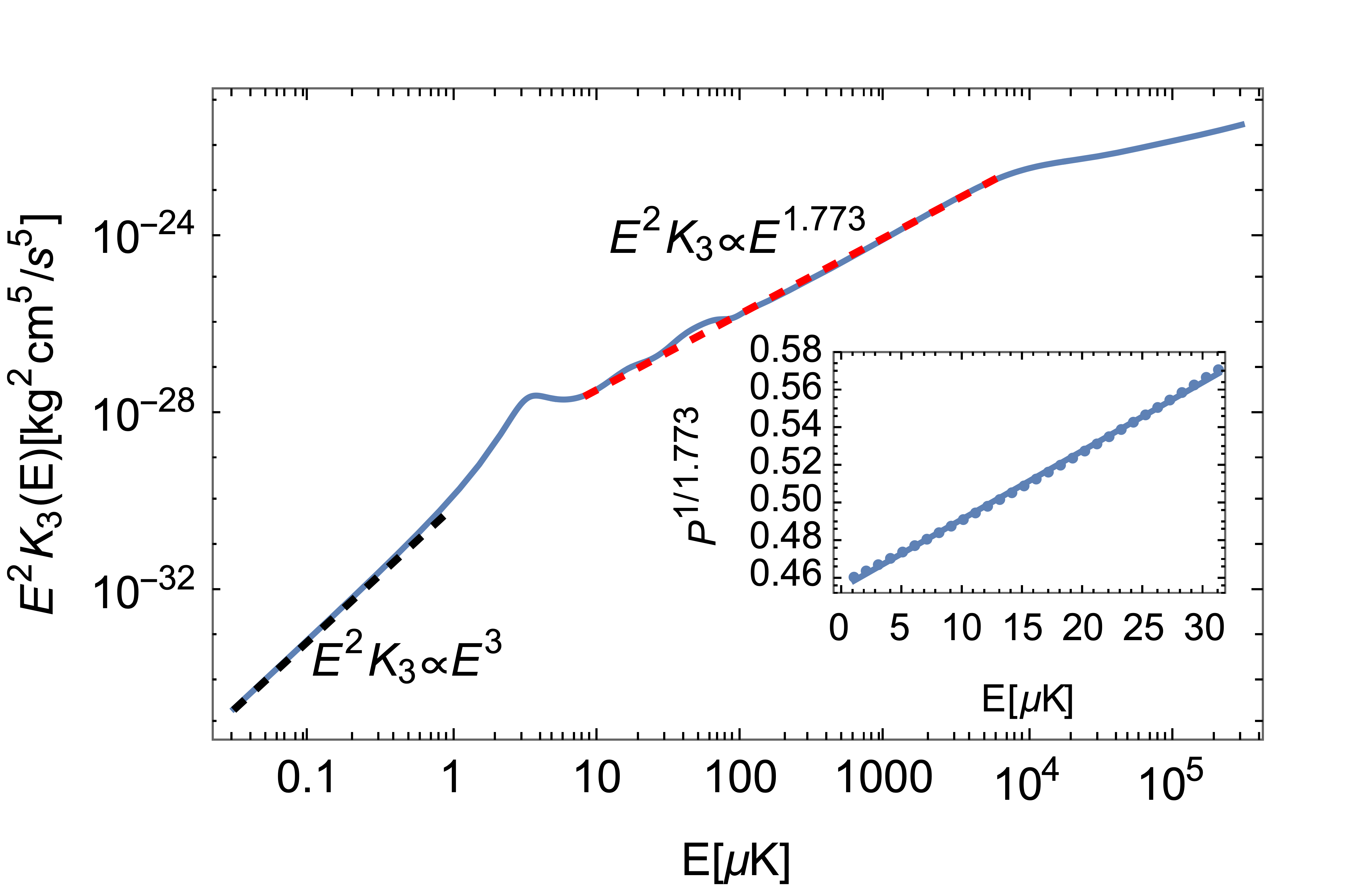}
\caption{(color online). Shown here is the rescaled recombination rate, $E^{2}K_{3}(E)$, as a function of the total energy $E$.  The interaction between opposite spin state fermions ($\uparrow\downarrow$) is set at $a_{s}=205\,r_{\text{vdW}}$ and the interaction between the two spin-up fermions ($\uparrow\uparrow$) is equal to $V_{p}=-1.8\,r_{\text{vdW}}^{3}$, corresponding to the three-body potential curves that are shown in Fig.(\ref{fig:pwave205}). The inset confirms that the power law for the WKB tunneling probability varies linearly with collision energy at very low energy, consistent with the expected Wigner threshold law for this symmetry.}
\label{fig:WKB}
\end{figure}

The Wigner threshold law for the three-body recombination in a low energy collision of two spin-up and one spin-down fermion with symmetry $L^{\Pi}=1^{-}$ is $K_{3}(E)\propto E$\cite{esry2001thresholdsupp}, for interactions not at unitarity. The power law of the three-body recombination depends on the number of identical particles and the system's angular momentum. However, when the two-body interaction is strong enough, i.e., the two-fermion interactions reside at the $s$-wave or $p$-wave unitary limit, the power law of the recombination becomes modified because there is a modified centrifugal barrier asymptotically.  In particular, the asymptotic $l_e$ value of the lowest 3-body continuum channel changes at unitarity, in a remnant of the Efimov effect\cite{chen2022efimovsupp,higgins2022threesupp}.  
Fig.(\ref{fig:WKB}) shows that in the very low energy limit, the threshold law of the three-body recombination rate into deep dimers does exhibit the expected linear dependence on the energy, $K_3(E)\propto E$ for collision energies from $0.03\,\mu K$ to $1\,\mu K$. However, the figure documents that the power law dependence of the recombination rate changes to $K_3(E)\propto E^{-0.227}$ when the temperature is in the range from $10 \,\mu  K$ to $10^{4}\, \mu K$. As a result, provided the $s$-wave scattering length is not too close to unitarity, the threshold law remains linear in $E$ in the low energy limit. The inset of Fig.(\ref{fig:WKB}) displays a WKB calculation  to analyze the threshold law of the three-body recombination loss rate. The WKB tunneling probability at the incident energy $E$ can be written as\cite{d2018fewsupp}
\begin{equation}
    P_{x \rightarrow y}^{(\nu)}=\exp \left\{-2 \int_{x}^{y}\sqrt{\frac{2\mu}{\hbar^{2}}\left[W_{\nu}(R)-E+\frac{1/4}{2\mu R^{2}}\hbar^{2}\right]}dR \right\}
\end{equation}
where the $\nu$ represents the $\nu$-th channel, $x$ and $y$ are the inner and outer classical turning points, respectively. Here $E$ is the incident collision energy and $\hbar^{2}(1/4)/(2\mu R^{2})$ is the semiclassical Langer correction\cite{berry1966semisupp}. The inset of Fig.(\ref{fig:WKB}) shows that the WKB probability has been raised to the power  $1/1.773$ and plotted versus the energy. 
The demonstrated proportionality between $P(E)$ and $E^{1.773}$ clarifies why the three-body recombination rate into deep dimers has this nonstandard (Efimov physics modified Wigner threshold law) in its near-threshold behavior at large $s$-wave scattering lengths.  

\onecolumngrid

\twocolumngrid

\subsection{Landau-Zener Probability}

\begin{figure}[ht]
\includegraphics[width=9cm]{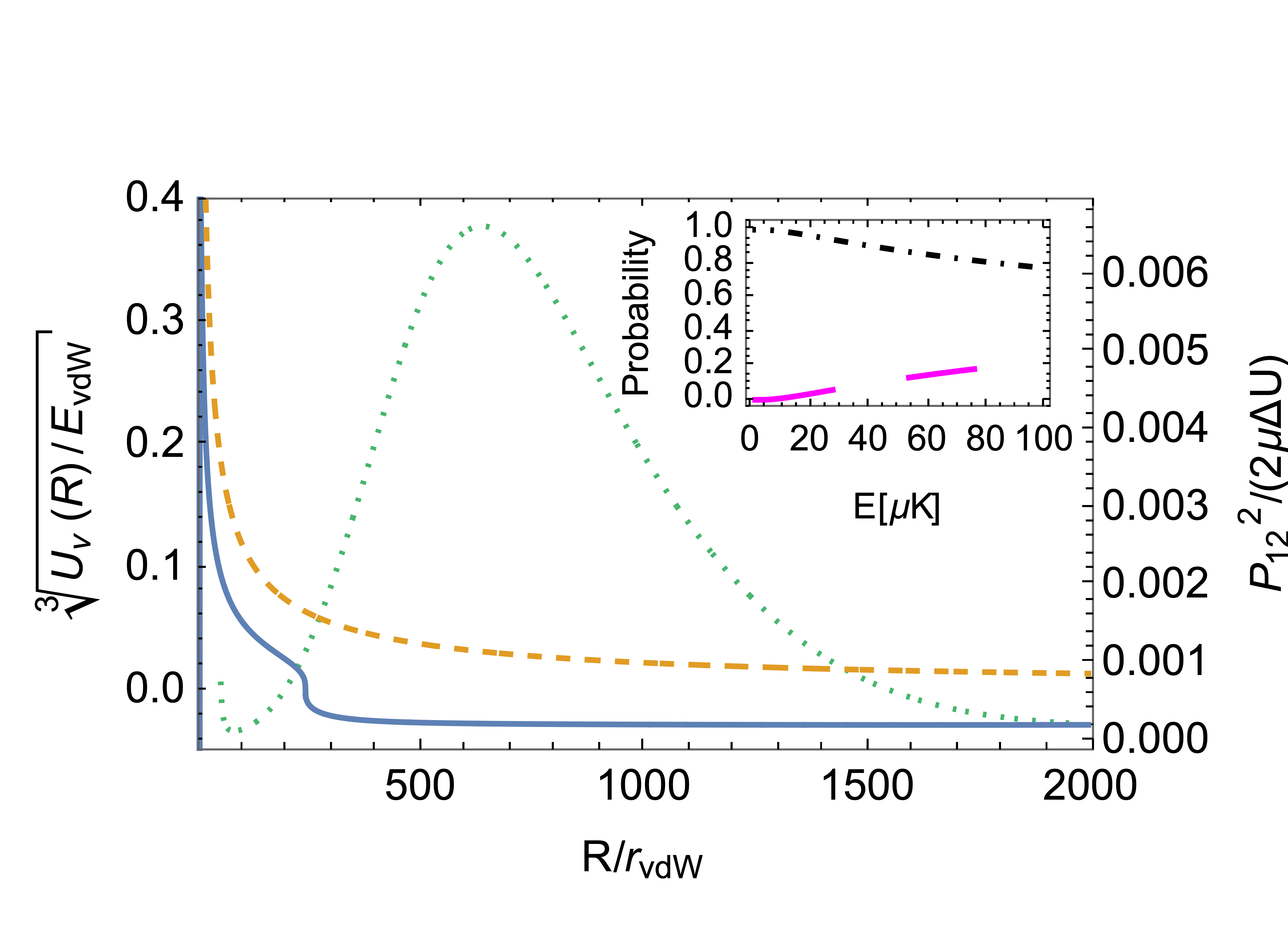}
\caption{(color online). The lowest two Born-Oppenheimer potential curves calculated near the first $s$- and $p$-wave poles are shown for symmetry $L^{\Pi}=1^{-}$. The interaction between the spin-up and spin-down fermions is set at $a_{s}=205\,r_{\text{vdW}}$ and between the two spin-up fermions at $V_{p}=-1.8\,r_{\text{vdW}}^{3}$. The solid line decreases below the three-body continuum threshold at around $R=240\,r_{\text{vdW}}$, while the dashed line is the lowest continuum channel potential curve, and the dotted line is a measure of the nonadiabatic coupling strength, showing a peak located near $R=619\,r_{\text{vdW}}$. The inset plots the Landau-Zener probability as a function of the collision energy. Specifically, the dash-dotted line (black) denotes the diabatic probability, and the long-dashed line (magenta) is the adiabatic probability.}
\label{fig:CubeU}
\end{figure}

\begin{figure}[ht]
\includegraphics[width=8cm]{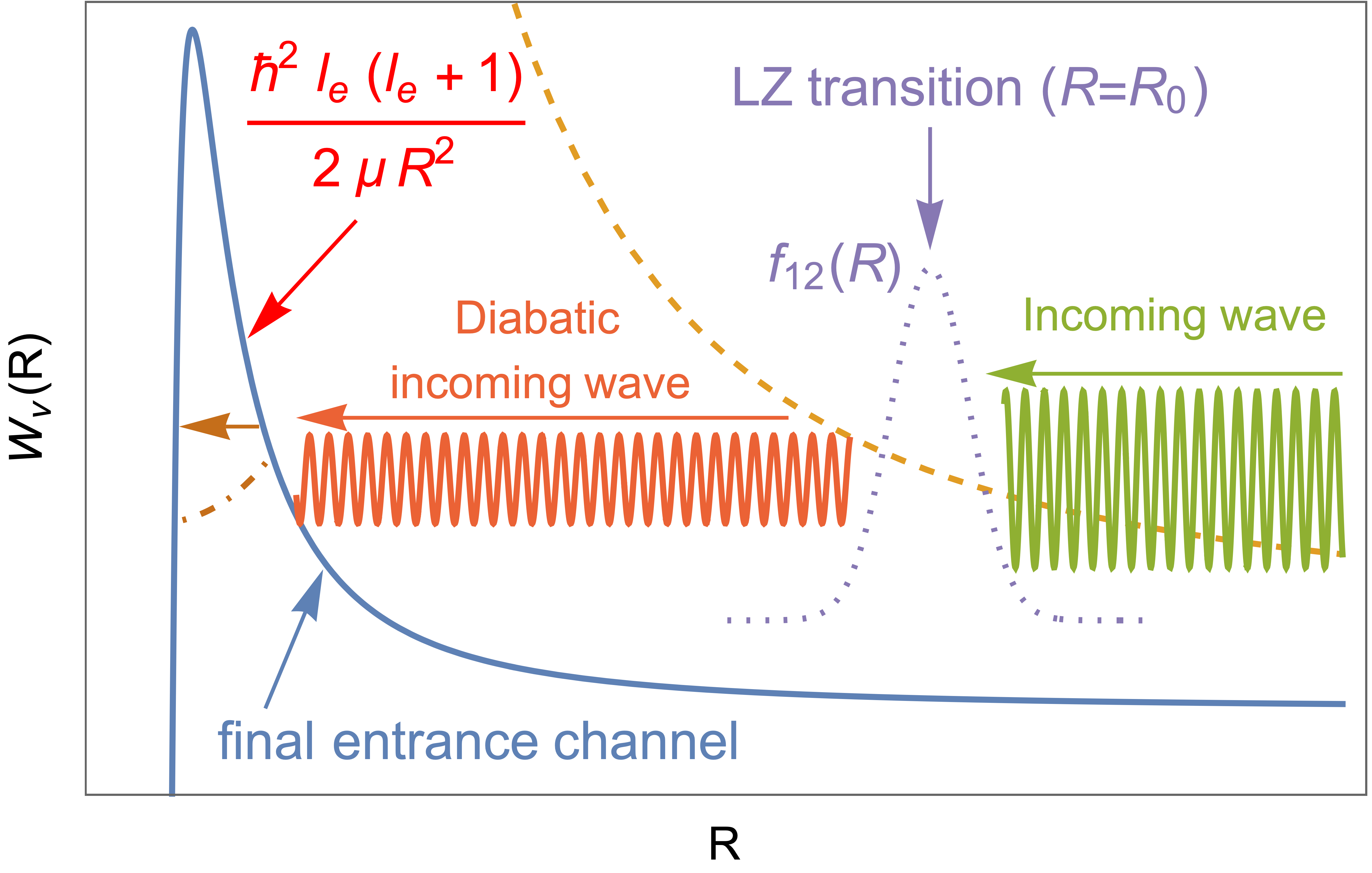}
\caption{(color online). Sketch of the Landau-Zener transition between the two lowest Born-Oppenheimer potential curves at the large positive $a_s$, where the $l_e$ value of the lower potential curve plays a key role in controlling the low energy threshold behavior of the three-body recombination rate. The solid curve is the lowest potential curve, which effectively acts as the three-body entrance channel at $R<R_0$, even though it goes below the three-body threshold and becomes an atom-dimer channel at very large $R\gg a_s$. The dashed curve plots the second lowest potential curve that becomes the second continuum channel at $R<R_0$ and which serves as the lowest 3-body continuum entrance channel at $R \gg R_0$. The dotted line shows the  measure of nonadiabatic coupling strength $f_{12}(R)$ and its peak position that is concentrated at $R=R_0$.
}
\label{fig:Sketch}
\end{figure}
If these adiabatic potential curves and their nonadiabatic couplings are computed out to very large hyperradii ($R \gg 10 a_s$), one could simply solve the coupled one-dimensional differential equations to obtain the full scattering matrix of the system.  That is impractical here for $a_s>0$ because there is a crucial avoided crossing between the two potentials sketched in Fig.(\ref{fig:Sketch}) out at huge hyperradii that are beyond our current computational capabilities.  However, we understand that the nonadiabatic coupling of those two channels can be approximately modeled as a Landau-Zener avoided crossing.
The nonadiabatic coupling matrices $P_{\nu\nu'}(R)$ and $Q_{\nu\nu'}(R)$ are defined as
\begin{align}
	&P_{\nu\nu'}(R)=\int d\Omega \Phi_{\nu}^{*}(R;\Omega)\frac{\partial}{\partial R}\Phi_{\nu'}(R;\Omega) \\
	&Q_{\nu\nu'}(R)=\int d\Omega \Phi_{\nu}^{*}(R;\Omega)\frac{\partial^{2}}{\partial R^{2}}\Phi_{\nu'}(R;\Omega)
\end{align} 
The following quantity serves as a useful measure of the nonadiabatic coupling strength between channels $\nu$ and $\nu'$: 
\begin{equation}
    f_{\nu\nu'}(R)=\frac{P_{\nu\nu'}(R)^{2}}{2\mu |U_{\nu}(R)-U_{\nu'}(R)|}
\end{equation}
Fig.\ref{fig:CubeU} plots the cube root of the lowest two Born-Oppenheimer potential curves as a function of hyperradius and compares it with the nonadiabatic coupling strength. The solid curve ($s$-wave atom-dimer channels) goes below the 3-body threshold at around $R=240\,r_{\text{vdW}}$ and converges asymptotically to the universal dimer energy appropriate to a large positive $a_s$.  The dashed curve is the lowest 3-body continuum channel that converges to 0 asymptotically, and the dotted curve represents the measure of nonadiabatic coupling strength $f_{12}(R)$ between these two channels. The peak of the nonadiabatic coupling means the three-body system has a high chance to diabatically transition as it moves inward from the continuum channel to the atom-dimer channel near the hyperradius $R\approx 3a_{s}$. The inset of Fig.\ref{fig:CubeU} shows the Landau-Zener transition probability versus the collision energy. The Landau-Zener probability to diabatically traverse an avoided crossing centered at the hyperradius $R_0$ can be represented as\cite{clark1979supp}:
\begin{equation}
    P_{diabatic}=\exp\left\{\frac{-2\pi U_{12}^{2}}{\hbar v |\frac{\partial}{\partial R}[U_{1}(R)-U_{2}(R)]|}\right\}
\end{equation}
where $U_{1}(R)$ and $U_{2}(R)$ are the two hyperspherical potential curves, $v$ is the incoming hyperradial velocity and $U_{12}=U_{1}(R_0)-U_{2}(R_0)$, and the slope difference in the denominator of the exponent is computed near but not exactly at $R_0$.  
In the inset, the Landau-Zener diabatic probability is seen to be close to unity in this low energy range, namely: $P_{diabatic}=0.977$, with the transition centered at $R=619\,r_{\text{vdW}}$ for the energy $E\approx 15.66\mu K$ in the range of energy and $s$-wave scattering length relevant for the experiment\cite{du2009inelasticsupp}. 
Note that the three-body recombination loss rate in Fig.3 has incorporated this diabatic probability factor from the Landau-Zener transition for the calculations performed at large positive $s$-wave scattering lengths. Fig.\ref{fig:Sketch} illustrates the mechanism of the Landau-Zener transition between two lowest potential curves and the $l_e$ value of lowest potential curve has significant effect on the threshold law of three-body recombination rate.  

One subtlety in our present calculations that requires explanation is the fact that for the large positive scattering lengths studied experimentally (in Fig.3 of the main article), there is a very weakly bound $s$-wave dimer of binding energy below $1 \mu K$. Recombination into that dimer is not expected to result in atom loss.  For positive $a_s$, therefore, the lowest energy {\it true} 3-body entrance channel is the dashed curve shown in Fig.\ref{fig:Sketch}.  However, in order to reach small hyperradii where recombination into deep dimers can occur, the system must reach the region left of the potential barrier of the solid (blue) potential curve in Fig.\ref{fig:Sketch}.  Because the two potential curves in the figure have an avoided crossing at $R_0 \approx 3 a_s$, this provides a dominant pathway for recombination into deep dimers that involves the incoming 3-body wave that transitions nonadiabatically into the solid potential curve labeled as the final entrance channel, and can then tunnel to the left of the potential barrier where recombination subsequently occurs.  

This sequence of steps that control deep dimer formation in such a collision has apparently not been described in previous studies of 3-body recombination.  Since $a_s \gg r_{\text{vdW}}$ in the range considered here, the solid (blue) potential curve in Fig.\ref{fig:Sketch} is the one that has its $l_e$ value affected strongly by Efimov physics, and for $R_{\rm barrier} \ll R \lesssim a_s$ it is reasonably well described by a centrifugal barrier whose coefficient is $l_e(l_e+1)$ with $l_e \approx 1.273$.  The tunneling amplitude through that barrier has an energy dependence described by an Efimov-modified Wigner threshold law factor $k^{l_e+1/2}$, as has been stressed in recent publications.\cite{chen2022efimovsupp, higgins2022threesupp}
Thus, the three-body recombination rate has the following energy dependence: $K_3\propto E^{l_{e}-3/2}=E^{-0.227}$, over the energy range from about 10 $\mu K$ up to 5 $mK$, and the WKB tunneling probability varies with energy as $P\propto E^{l_{e}+1/2}=E^{1.773}$, respectively (i.e., the exponent is $2-0.227$). Previous studies have shown how WKB tunneling under such a centrifugal barrier is one way of understanding the origin of the relevant Wigner threshold law for any given process\cite{Rau2000supp}. Therefore, the numerical thermally averaged 3-body loss rate into deep dimer formation can be described for large positive $a_s$ in terms of this pathway that involves the Landau-Zener transition probability followed by tunneling.  This reasoning is not relevant at large negative $a_s$ because in that case the solid (blue) curve of Fig.\ref{fig:Sketch} remains positive all the way out to $R\rightarrow \infty$ and consequently behaves just as a normal 3-body entrance channel, albeit with an Efimov-physics modified value of its centrifugal barrier equal to $l_e \approx 1.273$ out to $R\lesssim |a_s|$.  See Refs.\cite{higgins2022threesupp,chen2022efimovsupp}


\end{document}